\documentclass[showpacs,preprintnumbers,amsmath,amssymb,twocolumn,superscriptaddress,prb]{revtex4}
\usepackage{times}
\usepackage{graphicx}
\usepackage{amsfonts}
\usepackage{amsmath, amsthm, amssymb}
\usepackage{dsfont}
\usepackage{subfigure}

\newcommand{\e}[1]{\mathrm{e}^{#1}}

\newcommand{\etal}{\emph{et al. }}
\def\i{\mathrm{i}}

\begin{document}
\title[Identifying the odd-frequency superconducting state by a field-induced Josephson effect]
{Identifying the odd-frequency superconducting state by a field-induced Josephson effect}
\author{Jacob Linder}
\affiliation{Department of Physics, Norwegian University of
Science and Technology, N-7491 Trondheim, Norway}
\author{Takehito Yokoyama}
\affiliation{Department of Applied Physics, Nagoya University, Nagoya, 464-8603, Japan}
\author{Asle Sudb{\o}}
\affiliation{Department of Physics, Norwegian University of
Science and Technology, N-7491 Trondheim, Norway}

\date{Received \today}
\begin{abstract}
Superconducting order parameters that are odd under exchange of time-coordinates of the electrons constituting a 
Cooper-pair, are potentially of great importance both conceptually and technologically. Recent experiments report that
such an odd-frequency superconducting  {\it bulk} state may be realized in certain heavy-fermion compounds. While 
the Josephson current normally only flows between superconductors with the same symmetries with respect 
to frequency, we demonstrate that an exchange field may induce a current between diffusive even- and 
odd-frequency superconductors. This suggests a way to identify the possible existence of bulk odd-frequency 
superconductors.
\end{abstract}

\maketitle

\section{Introduction}
The prevalent symmetry in known superconductors may be described as odd under exchange of spin coordinates, 
and even under an exchange of spatial coordinates or an exchange of time coordinates of the electrons 
constituting the Cooper pair. The latter condition corresponds to a sign change of frequency, after Fourier-transforming the time variables to a frequency representation. This symmetry may compactly be 
expressed as an even-frequency singlet even-parity superconducting state (hereafter referred to as the 
even-frequency state). 
\par
However, other types of pairing are also permitted. Among these is the so-called odd-frequency pairing 
state \cite{berezinskii}. Such a state is potentially of great importance, 
both from a conceptual as well as a technological point of view. From a conceptual point of view
phase transitions involving pairing of fermions form centerpieces of physics in such widely disparate 
sub-disciplines as cosmology, astrophysics, physics of condensed matter, physics of extremely 
dilute  ultra-cold atomic gases, and physics of extremely compressed quantum liquids. Extending the possible 
pairing states compatible with the Pauli-principle will likely have impact on all these disciplines, and 
hence deserves attention. From a technological point of view, an odd-frequency pairing state would be a 
candidate for a robust superconducting pairing state capable of coexisting with ferromagnetism \cite{bergeretRMP}.  Such a 
material would be extremely important, since it would combine two functionalities of major importance, 
namely magnetism and superconductivity. 
\par
The hope of experimentally detecting this type of pairing was raised by the prediction that an odd-frequency 
triplet even-parity (hereafter simply denoted odd-frequency) pairing should be induced by the proximity 
effect in diffusive ferromagnet/conventional superconductor junctions \cite{bergeretPRL}. Recently, 
odd-frequency states due to the proximity effect have been confirmed in experiments of diffusive 
ferromagnet/conventional superconductor junctions \cite{Kaizer}. To date, no bulk odd-frequency 
superconductor has been unambigously identified. 
\par
\begin{figure}[h!]
\centering
\resizebox{0.48\textwidth}{!}{
\includegraphics{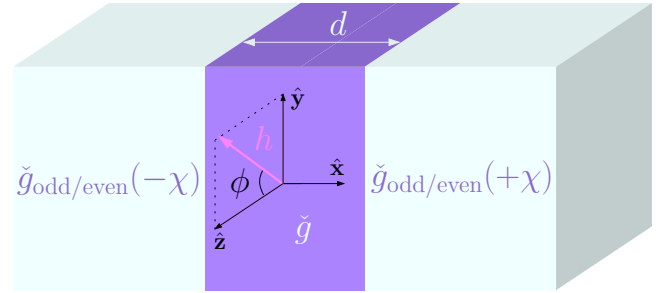}}
\caption{A diffusive metal which may be either normal or ferromagnetic (with exchange field $h$) of width $d$ separates 
two superconductors with a phase difference $2\chi$, thus constituting a Josephson current ($I$ is the current source). The individual superconductors have either an even-frequency or 
odd frequency bulk symmetry. The exchange field may have any orientation in the $yz$-plane, characterized by the angle $\phi$.}
\label{fig:model}
\end{figure}
Currently, the heavy-fermion compounds CeCu$_2$Si$_2$, and CeRhIn$_5$ seem to be the most promising candidates for the realization of a bulk odd-frequency state \cite{fuseya}. It has also recently been argued \cite{johannes} that this type of pairing could be realized in Na$_x$CoO$_2$, motivated by band-structure calculations and the robustness of its superconductivity against impurities.
However, the experimental reports on the Knight-shift data have shown evidence of both singlet \cite{kobayashi} and triplet \cite{ihara} pairing. To resolve the pairing issue, it would be desirable to make clear-cut theoretical predictions for experimentally 
measurable quantities that may distinguish the odd-frequency symmetry from the conventional even-frequency symmetry. 
Motivated by this, the conductance of a diffusive normal metal (N) /odd-frequency junction was recently studied \cite{fominov}. 
\par
It is well-known that the Josephson current between superconductors with different symmetries is in general inhibited 
\cite{abrahams}. However, it was very recently shown that in the clean limit, such a Josephson coupling may be established 
between such superconductors by means of surface-induced pairing components of different symmetry than the bulk state in 
each superconductor \cite{tanakaPRLNEW}. In the dirty limit, it has been demonstrated that in the case of an $s$-wave even- or 
odd-frequency bulk superconductor, the proximity-induced pairing component in the N will have the same symmetry 
\cite{tanakaPRL07}. These results are valid in the absence of an exchange field. Due to the above mentioned properties, 
one would consequently expect that both the even- and odd-frequency symmetries are induced in the normal part of an 
even-frequency/N/odd-frequency junction. While this is true, the Josephson current is found to vanish in such a setup. 
If the N is replaced with a diffusive ferromagnet (F), the pairing components in the F will have the same symmetries 
as in the N case. Surprisingly, the Josephson current is \textit{not} absent in this case. 
\par
In this paper, we report that a Josephson current may flow in a diffusive even-frequency/F/odd-frequency junction, and 
argue that this should serve as a \textit{smoking gun} to reveal the odd-frequency symmetry in a bulk superconductor. We 
also study the dependence of the Josephson current on the temperature and width of the F, and show that $0$-$\pi$ transitions 
take place. In the following, we will use boldface notation for 3-vectors, $\check{\ldots}$ for $8\times8$ matrices, 
$\hat{\ldots}$ for $4\times4$ matrices, and $\underline{\ldots}$ for $2\times2$ matrices.

\section{Theoretical framework}
In order to address this problem, we employ the quasiclassical theory of superconductivity using the Keldysh formalism. The 
quasiclassical Green's functions may be divided into an advanced (A), retarded (R), and Keldysh (K) component, each of which 
has a $4\times4$ matrix structure in the combined particle-hole and spin space. In thermal equilbrium, it suffices to consider 
the retarded component since the advanced component is obtained by 
$\hat{g}^\text{A} = -(\hat{\rho}_3 \hat{g}^\text{R} \hat{\rho}_3)^\dag$, 
while the Keldysh component is given by 
$\hat{g}^\text{K} = (\hat{g}^\text{R} - \hat{g}^\text{A})\tanh(\beta\varepsilon/2)$, 
where $\beta$ is inverse temperature. Pauli-matrices in particle-hole$\times$spin (Nambu) space are denoted as $\hat{\rho}_i$, 
while Pauli-matrices in spin-space are written as $\underline{\tau}_i$, $i=1,2,3$. As shown in Ref. \cite{tanakaPRL07}, the 
symmetry of the anomalous Green's function induced in the N through the proximity effect by an odd-frequency superconductor 
is also odd-frequency, and similarly for an even-frequency superconductor. We may write the retarded Green's function in the 
superconductors as
\begin{align}
\hat{g}^\text{R}_ \text{odd/even} = \begin{pmatrix}
c\underline{1} & s\e{\pm\i\chi}\underline{\tau_1} / s\e{\pm\i\chi}\i\underline{\tau_2}  \\
-s\e{\mp\i\chi}\underline{\tau_1} / s\e{\mp\i\chi}\i\underline{\tau_2} & -c\underline{1}\\
\end{pmatrix}
\end{align}
for an odd-frequency triplet even parity $s$-wave symmetry / even-frequency singlet even parity $s$-wave symmetry. Above, we have defined 
\begin{align}
c\equiv\text{cosh}(\theta),\; s\equiv\text{sinh}(\theta),\; \theta \equiv \text{atanh}[\Delta(\varepsilon)/\varepsilon],
\end{align}
where $\varepsilon$ is the quasiparticle energy measured from Fermi level while $\chi$ is the superconducting phase associated with the broken U(1) symmetry. The $\pm$-sign refers to the right and left side of the F (Fig. \ref{fig:model}). The retarded Green's function in the F satisfies the Usadel equation \cite{usadel} 
\begin{align}
&D\nabla (\hat{g}^\text{R}\nabla\hat{g}^\text{R}) + \i[\varepsilon\hat{\tau}_3 + \hat{M}, \hat{g}^\text{R}] = 0, \notag\\
\hat{M} &= h\begin{pmatrix}
\cos\phi\underline{\tau_3} + \sin\phi\underline{\tau_2} & \underline{0} \\
\underline{0} & (\cos\phi\underline{\tau_3} + \sin\phi\underline{\tau_2})^\mathcal{T} \\
\end{pmatrix},
\label{Usadel}
\end{align}
where $D$ is the diffusion constant, $h$ is the exchange field, and $\mathcal{T}$ denotes the matrix transpose. An analytical solution of this equation is permissable when it may be linearized, which corresponds to a weak proximity effect. This may be obtained in two limiting cases: 1) if the barriers have low transparency or 2) if the transmission is perfect (ideal interfaces) and the temperature in the superconducting reservoir is close to $T_c$, such that $|\Delta|$ is small. We will here mainly focus on the low transparency case \cite{boundary_tanaka} since it might be difficult to experimentally realize small and highly transparent
junctions for observing the Josephson current. In this case, we make use of the standard Kupryianov-Lukichev boundary conditions \cite{kupluk}
\begin{align}
2\gamma d (\hat{g}\partial_x \hat{g})  =  \pm [\hat{g}, \hat{g}_\text{R(L)}(\pm\chi)]|_{x=d,0},
\end{align}
where $\gamma$ is a measure of the barrier strength and $\hat{g}_\text{R(L)}$ denotes the Green's function in the superconductor on the right ($x=d$) or left ($x=0$) side of the F, respectively. Defining the vector anomalous Green's function
\begin{equation}
\mathbf{f} = [f_\downarrow-f_\uparrow, -\i(f_\uparrow+f_\downarrow), 2f_\text{t}]/2.
\end{equation}
and the matrix anomalous Green's function in spin-space,
\begin{equation}
\underline{f} = (f_\text{s} + \mathbf{f}\cdot\underline{\boldsymbol{\tau}})\i\underline{\tau_2}  = \begin{pmatrix}
f_\uparrow & f_\text{t}+f_\text{s}\\
f_\text{t}-f_\text{s} & f_\downarrow\\
\end{pmatrix},
\end{equation}
the linearized Green's function in the F reads in total
\begin{align}\label{eq:ff}
\hat{g}^\text{R} = \begin{pmatrix}
\underline{1} & \underline{f}(\varepsilon) \\
-[\underline{f}(-\varepsilon)]^* & -\underline{1}\\
\end{pmatrix}.
\end{align}
Note that Eq. (\ref{eq:ff}) contains both equal-spin and opposite-spin pairing triplet components in general. In the special cases of $\phi=0$ and $\phi=\pi$, the equal-spin pairing components $f_\sigma$ $(\sigma=\uparrow,\downarrow)$ vanish.

\section{Results}
We now provide the analytical results for the Josephson current in an even-frequency/F/odd-frequency junction. To begin with, we will consider the case $\phi=0$ to emphasize our main result: namely a \textit{field-induced} Josephson effect. We will then investigate the effect of a change in orientation of the magnetization in the ferromagnetic layer. Changing the orientation, i.e. $\phi$, in an even-frequency/F/even-frequency junction has no effect since the order parameters in the superconductors in that case are isotropic. However, for an odd-frequency triplet superconductor the order parameter has a direction in spin-space. We here consider an opposite-spin pairing $S_z=0$ order parameter without loss of generality, and then proceed to vary the orientation of the exchange field. 

\subsection{Field-induced Josephson current}
For $\phi=0$, one readily finds that $f_\sigma=0$. The linearized version of equation Eq. (\ref{Usadel}) is then given by 
\begin{align}
D\partial_x^2f_\pm + 2\i(\varepsilon\pm h) f_\pm=0,
\end{align}
with the general solution 
\begin{align}
f_\pm(x) &= A_\pm\e{\i k_\pm x} + B_\pm\e{-\i k_\pm x},\notag\\
k_\pm &= \sqrt{2\i(\varepsilon\pm h)/D},
\end{align}
with the definition $f_\pm = f_\text{t}\pm f_\text{s}$. Employing the boundary conditions, one finds that the anomalous Green's function reads
\begin{align}\label{eq:f}
f_{\pm} &= \frac{B_{\pm}(c_\text{L} + \rho_\pm) \mp s_\text{L}\e{-\i\chi}}{(\rho_\pm - c_\text{L})\e{-\i k_\pm x}} + B_{\pm}\e{-\i k_\pm x},
%f_{2,\pm} &= (\pm s_\text{L}\e{-\i\chi} - B_{2,\pm})\e{\i k_\pm x} + B_{2,\pm}\e{-\i k_\pm x},
\end{align} 
where we have introduced the auxiliary quantity
\begin{align}\label{eq:B}
B_{\pm} &= [s_\text{R}\e{\i\chi} \pm s_\text{L}\e{\i(k_\pm d-\chi)}(\rho_\pm + c_\text{R})/(\rho_\pm-c_\text{L})]\notag\\
&\times[\e{\i k_\pm d}(\rho_\pm + c_\text{L})(\rho_\pm + c_\text{R})/(\rho_\pm - c_\text{L}) \notag\\
&- \e{-\i k_\pm d}(\rho_\pm - c_R)]^{-1}.
%B_{2,\pm} &= [\pm s_\text{L}\e{\i(k_\pm d -\chi)} - s_\text{R}\e{\i\chi}][2\i\sin(k_\pm d)]^{-1},
\end{align}
Moreover, we have introduced $\rho_\pm = \gamma d\i k_\pm$ while the subscripts 'L' and 'R' denote the left and right superconductor for the coefficients $c$ and $s$.
Once $\hat{g}^\text{R}$ has been obtained, the Josephson current may be calculated by the formula 
\begin{align}\label{eq:jos1}
\mathbf{j}(x) &= -(N_\text{F}eD\hat{\mathbf{x}}/4)\int\text{d}\varepsilon \text{Tr}\{\hat{\rho}_3 (\check{g}\partial_x\check{g})^\text{K}\}\notag\\
&= -(N_\text{F}eD\hat{\mathbf{x}}/2) \int^\infty_{-\infty} \text{d}\varepsilon \text{Re}\{M_+(\varepsilon)+M_-(\varepsilon)\}\notag\\
&\hspace{1.3in}\times\tanh(\beta\varepsilon/2),
\end{align}
with the definition
\begin{align}\label{eq:jos}
M_{\pm}(\varepsilon)=[f_\pm(-\varepsilon)]^*\partial_x f_\mp(\varepsilon) - f_\pm(\varepsilon)\partial_x[f_\mp(-\varepsilon)]^*.
\end{align}
The normalized current density is defined as 
\begin{align}\label{eq:critical}
I(\chi)/I_0 = 4|\mathbf{j}(x,\chi)|/(N_\text{F}eD\Delta_0^2),
\end{align}
which is independent of $x$ for $x\in[0,d]$, and the critical current is given by $I_\mathrm{c} = I(\frac{\pi}{4})$. 
\par
At this point, we are in a position to compare the results for the even- and odd-frequency case against each other, to investigate how the different symmetry properties alter the Josephson current. Similarly to Ref. \cite{tanakaPRL07}, we will model the odd-frequency gap by 
\begin{align}
\Delta(\varepsilon) = \varepsilon/[1+(\varepsilon/\Delta_0)^n],
\end{align}
 where $n=2,4,8,..$, which exhibits the low-energy behaviour considered in Ref. \cite{fominov}. We choose specifically $n=2$, and underline that none of our qualitative conclusions are altered by choosing  $n=4,8$. It is natural to begin looking for signatures of the odd-frequency symmetry as probed by the Josephson current in the simplest case of an even-frequency/N/odd-frequency junction. The proximity-induced anomalous Green's function in the N will have a contribution from both the even- and odd-frequency symmetry \cite{tanakaPRL07}. However, we find that \textit{no Josephson current may flow} in such a setup. 
\par
This may be understood by considering the boundary conditions at each interface (see Fig. \ref{fig:anomalous}). At the even-frequency/N interface, the odd-frequency triplet component of the proximity-induced Green's function is absent since penetration into the even-frequency superconductor is prohibited. Similarly, at the N/odd-frequency interface, the even-frequency singlet component of the Green's function vanishes for the same reason. 
\par
Therefore, the Josephson current is zero in this type of junctions since the current-carrying Green's function in the N induced by the left superconductor is absent at the right superconductor, and vice versa. This is a direct result of the different symmetries of the even- and odd-frequency superconductors. The Josephson current can be divided into the individual contributions from $f_\text{s}$ and $f_\text{t}$ (cross terms vanishes), and for each component the 
coherence is lost since the even- and odd-frequency pairings cannot reach the opposite 
interface. 
\par
Analytically, one can confirm from Eq. (\ref{eq:f}) and (\ref{eq:B}) that the Green's functions providing the critical current (at $\chi=\pi/4$) satisfy $f_\pm(\varepsilon) = \i f_\pm^*(-\varepsilon)$ for any $x$ in the even-frequency/N/odd-frequency case, which upon insertion in Eq. (\ref{eq:jos}) yields $\mathbf{j}(x)=0$. In even-frequency/N/even-frequency and odd-frequency/N/odd-frequency junctions, one can in a similar manner confirm that $A(\varepsilon) = A^*(-\varepsilon)\e{-\i kd}$ and $B(\varepsilon) = B^*(-\varepsilon)\e{\i kd}$, which leads to a finite value of the Josephson current. This may be seen for instance at $x=d/2$ upon substitution into Eq. (\ref{eq:jos}), where $f_\pm(\varepsilon) = f_\pm^*(-\varepsilon)$ and $\partial_x f_\pm(\varepsilon) = -\partial_xf^*_\pm(-\varepsilon)$.
\par
\begin{widetext}
\text{ }\\
\begin{figure}[h!]
\centering
\resizebox{0.7\textwidth}{!}{
\includegraphics{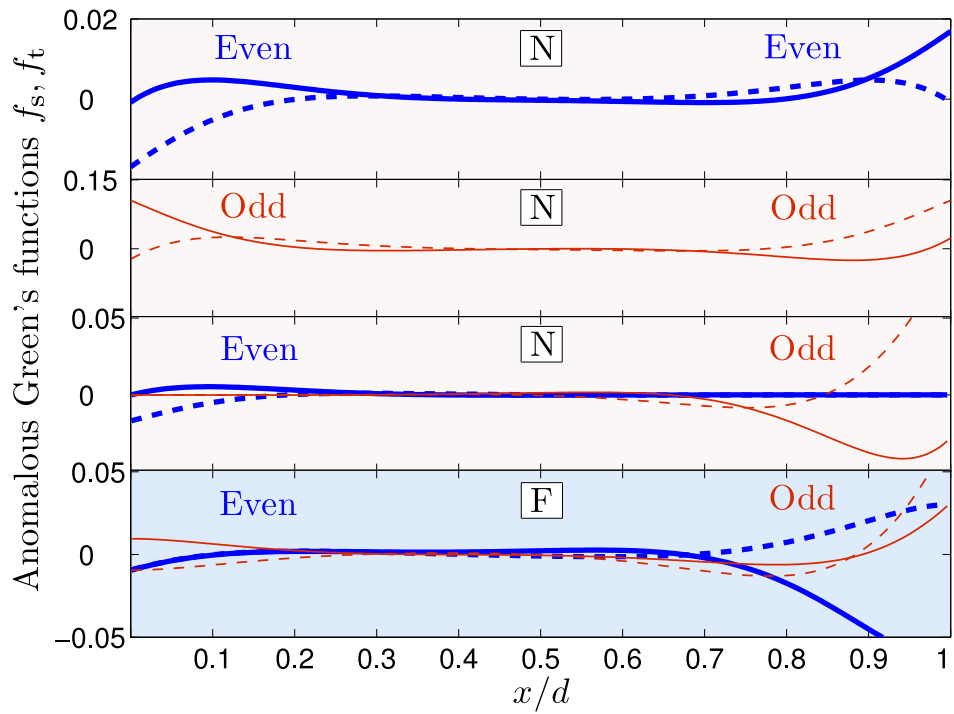}}
\caption{(Color online) Plot of the singlet $f_\text{s}$ (thick, blue lines) and triplet $f_\text{t}$ (thin, red lines) anomalous Green's function induced in a N (upper panels) and F (lower panel) in proximity with an even-frequency and/or odd-frequency superconductors, using $\chi=\pi/4$ and $\varepsilon/\Delta_0 = 0.1$. The solid lines are the real parts of the anomalous Green's functions, while the dashed lines are the imaginary parts. To ensure that $f_\text{s,t}\ll1$, we take $d/\xi=50$ in the N case and $d/\xi = 2$, $h/\Delta_0 = 50$ in the F case. Here, $\xi=\sqrt{D/(2\pi T_c)}$ is the superconducting coherence length. While a Josephson current flows when the symmetries of the superconductors are equal in the N case, it vanishes when the superconductors have \textit{different} symmetries. This is because the triplet and singlet components are absent at the interface of the even- and odd-frequency superconductor, respectively. Replacing the N with a F establishes a Josephson coupling between superconductors with different symmetries.}
\label{fig:anomalous}
\end{figure}
\end{widetext}
If we instead consider an even-frequency/F/odd-frequency junction, the proximity-induced anomalous Green's function in the F will have 
the same symmetries as in the N case. Remarkably, we find that in this case, however, \textit{a Josephson current 
is allowed to flow} through the system. The reason for why a Josephson current is present when the N is replaced 
with a F, is that the exchange field allows for both the singlet and triplet components to be induced throughout the F due to the mixing of singlet and triplet components by breaking the symmetry in spin space, regardless of the internal symmetry of the superconductors \cite{TK,yokoyamaPRB07} as shown in Fig. \ref{fig:anomalous}. As a result, we suggest that as a way to unambigously identify an odd-frequency superconductor, a coupling through a N to an ordinary even-frequency superconductor should not 
yield any Josephson current while replacing the N with an F should allow for the current to flow. One could argue that this is precisely the case also for even-frequency triplet odd-parity superconductors \cite{yokoyamaPRB07}. This symmetry is nevertheless easily distinguished from an odd-frequency $s$-wave 
symmetry since the former is highly sensitive to impurities 
while the latter does not suffer from this drawback. 
\par
\begin{figure}[h!]
\centering
\resizebox{0.48\textwidth}{!}{
\includegraphics{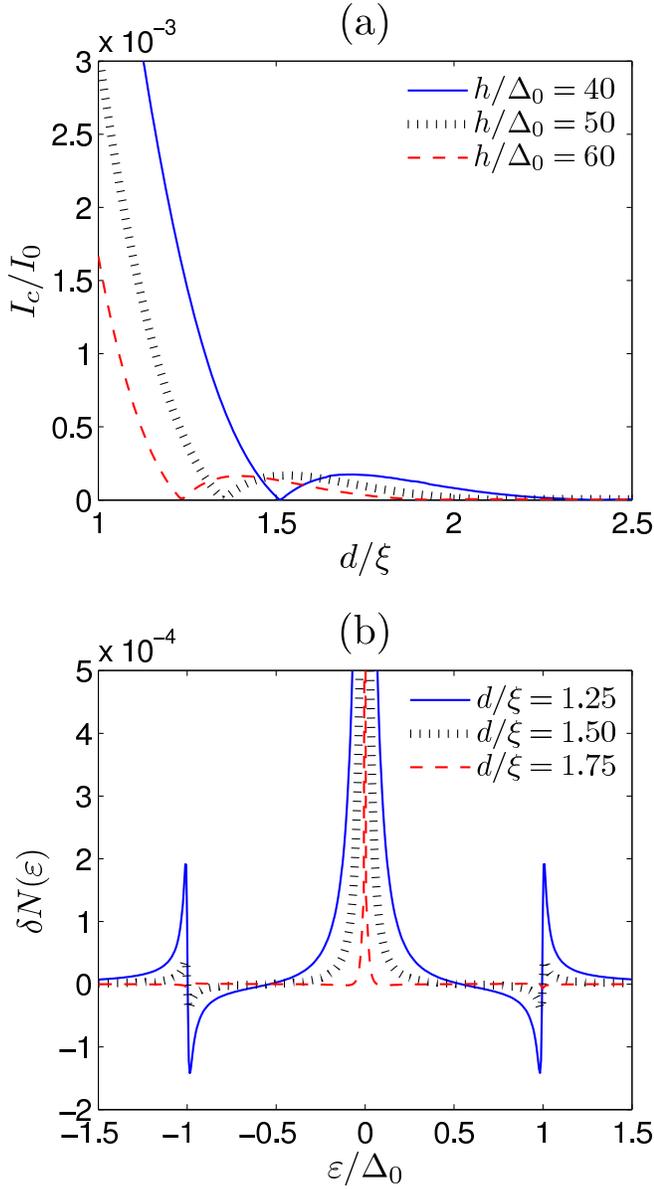}}
\caption{(Color online) a) The critical current $(\chi=\pi/4)$ as a function of $d/\xi$ for $T/T_\mathrm{c} = 0.001$.  b) Deviation from the LDOS at $x=d/2$ for an even-frequency/F/odd-frequency junction with $\chi=\pi/4$ and $h/\Delta_0 = 50$. The zero-energy peak is clearly discernible, originating with the odd-frequency pairs. }
\label{fig:Fig1}
\end{figure}

We have investigated the behaviour of the critical current when coupling an even- and odd-frequency superconductor through a F, and the results are shown in Fig. \ref{fig:Fig1} for a representative choice of parameters. To make the numerical calculations stable, we have added a small imaginary number \cite{Dynes}
in the quasiparticle energy, $\varepsilon\to\varepsilon+\i\delta$ with $\delta=0.01\Delta_0$, and we fix $\gamma=5$ (also for Fig. \ref{fig:anomalous}).
One observes the well-known $0$-$\pi$ transitions \cite{Bulaevskii} upon increasing with the width $d$ of the F [Fig. \ref{fig:Fig1}a)]. Similarly, we also find that $0$-$\pi$ oscillations occur as a function of temperature. The current-phase relationship is sinusoidal as usual in the linearized treatment. The width of the junction is measured in units of the superconducting coherence length \footnote{Some authors also use the definition $\xi=\sqrt{D/\Delta_0}$, which is of the same order of magnitude.} $\xi=\sqrt{D/(2\pi T_c)}$.
\par
As a further probe of the odd-frequency symmetry in a bulk superconductor, we investigate the local density of states (LDOS) in the F. 
The LDOS is altered from its normal state value due to the proximity-induced anomalous Green's function in the F. Consider Fig. \ref{fig:Fig1}b) for a plot of the deviation $\delta N$ from the normal state LDOS in an even-frequency/F/odd-frequency junction. The deviation is given by the formula
\begin{align}\label{eq:dos}
\delta N(\varepsilon) = \sum_{\sigma=\pm} (\text{Re}\{\sqrt{1 + f_\sigma(\varepsilon)[f_{-\sigma}(-\varepsilon)]^*}\} - 1)/2
\end{align}
under the assumption of a weak proximity effect.
As is seen, the LDOS is enhanced at $\varepsilon=0$ due to the presence of odd-frequency pairs \cite{tanakaPRL07,Braude}, and the usual peak arises at $\varepsilon=\Delta_0$. While the odd-frequency/F/odd-frequency case exhibits the first property, and the even-frequency/F/even-frequency case the latter, the even-frequency/F/odd-frequency junction is characterized by the fact that both of these features appear in $\delta N$. This could serve as an identifier of the odd-frequency symmetry in conjunction with the other properties we have analyzed here. 
\par
\begin{figure}[h!]
\centering
\resizebox{0.48\textwidth}{!}{
\includegraphics{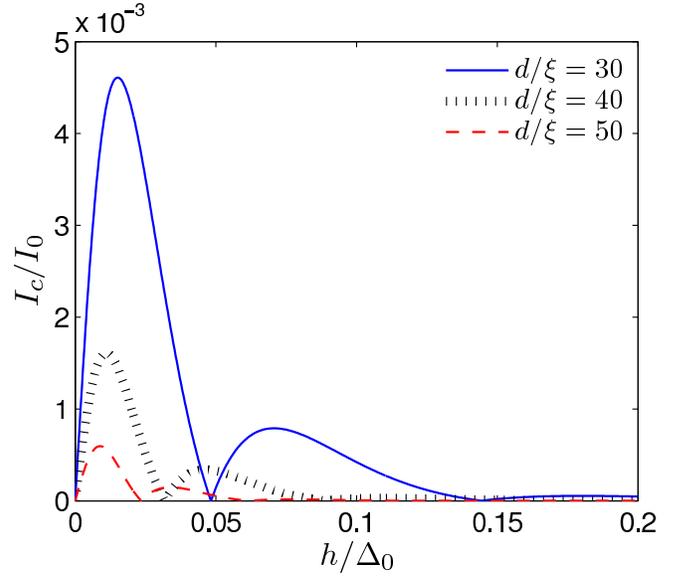}}
\caption{(Color online) Field dependence of the critical current for several values of $d/\xi$ with $T/T_c=0.001$.}
\label{fig:current_h}
\end{figure}

It is interesting to observe the field-dependence of the critical current in an even-frequency/F/odd-frequency junction, shown in Fig. \ref{fig:current_h}. To gain access to the regime of a very weak or absent exchange field, we must choose the width $d$ sufficiently large to ensure a weak proximity effect. In Fig. \ref{fig:current_h}, we plot the current as a function of $h$ for several values of $d/\xi$. One observes that the critical current goes exactly to zero at $h=0$, while a current is induced for non-zero values of $h$. A maximum peak appears for very weak exchange fields, and the critical current oscillates with increasing field-strength. Although the short-junction regime is not accessible for very weak exchange fields due to the linearized treatment, it follows from our analytical expressions that the current is absent for any choice of $d$ as long as $h=0$. 

\subsection{Case with $\phi\neq 0$}
We now proceed to consider the effect of rotating the exchange field in the ferromagnetic region. In effect, we allow for $\phi\neq0$. Now, the equal-spin pairing components $f_\sigma$ are in general non-zero, and we find the following four coupled, linearized Usadel equations:
\begin{align}
D \partial_x^2 f_\pm + 2\i(\varepsilon+h\cos\phi) f_\pm \pm h\sin\phi (f_\uparrow+f_\downarrow) = 0,\notag\\
D\partial_x^2 f_\sigma + 2\i\varepsilon f_\sigma - 2h\sin\phi f_s = 0,\; \sigma=\uparrow,\downarrow.
\end{align}
The general solution for the anomalous Green's functions may be written compactly as:
\begin{align}\label{eq:case2a}
f_\pm(x) &= A(x)(a_\uparrow+a_\downarrow) + B(x)(b_\uparrow+b_\downarrow) \notag\\
&+ c_5 R_+^\pm(x) + c_6 P^\pm_+(x) + c_7R_-^\pm(x) + c_8 P^\pm_-(x),\notag\\
\text{ }\notag\\
f_\sigma &= a_\sigma\cos L_0x + b_\sigma \sin L_0x + c_5(\i\sin L_+x\sin\phi) \notag\\
&+ c_6(\i\cos L_+x\sin\phi) + c_7(-\i\sin L_-x\sin\phi) \notag\\
&+ c_8(-\i\cos L_-x\sin\phi),\; \sigma=\uparrow,\downarrow.
\end{align}
In the above, $\{a_\sigma,b_\sigma,c_5,c_6,c_7,c_8\}$ are unknown coefficients to be determined 
from the boundary conditions in the problem. Also, we have introduced the following auxiliary 
quantities:
\begin{align}\label{eq:case2b}
A(x) &= \frac{\i\cos L_0x\sin\phi}{2\cos\phi},\; B(x) = \frac{\i\sin L_0x\sin\phi}{2\cos\phi},\notag\\
R_\pm^{\alpha}(x) &= \sin L_\pm x (\pm\cos\phi + \alpha),\; \alpha=\pm1\notag\\
P_\pm^{\alpha}(x) &= \cos L_\pm x(\pm\cos\phi + \alpha),\; \alpha=\pm1\notag\\
L_0 &= (1+\i)\sqrt{\frac{\varepsilon}{D}},\; L_\pm = (\i\pm1)\sqrt{\frac{h\pm\varepsilon}{D}}.
\end{align}
When the left superconductor has an even-frequency symmetry and the right superconductor has an odd-frequency symmetry, the boundary conditions yield at $x=0$:
\begin{align}
2\gamma d \partial_x f_\pm &= 2f_\pm c_L \mp 2s_L \e{-\i\chi},\notag\\
2\gamma d \partial_x f_\sigma &= 2f_\sigma c_L,
\end{align}
while at $x=d$ one obtains
\begin{align}
2\gamma d \partial_x f_\pm &= 2s_R \e{\i\chi} - 2f_\pm c_R ,\notag\\
2\gamma d \partial_x f_\sigma &= -2f_\sigma c_R.
\end{align}
The presence of equal-spin pairing components slightly modified the expressions Eq. (\ref{eq:jos1}) and Eq. (\ref{eq:dos}) for the density of states and the Josephson current, respectively. We now obtain
\begin{align}
\delta N(\varepsilon) &= \sum_\sigma [N_\sigma(\varepsilon) - 1]/2,\notag\\
N_\sigma(\varepsilon) &= \text{Re}\Big\{ \{1 + f_\sigma(\varepsilon)f_\sigma^*(-\varepsilon) + [f_t(\varepsilon) \notag\\
&+ \sigma f_s(\varepsilon)][f_t^*(-\varepsilon) - \sigma f_s^*(-\varepsilon)] \}^{1/2} \Big\}
\end{align}
for the density of states, while the Josephson current is calculated according to
\begin{align}\label{eq:jos45}
\mathbf{j}(x) &= -(N_\text{F}eD\hat{\mathbf{x}}/2) \int^\infty_{-\infty} \text{d}\varepsilon \text{Re}\{M_+(\varepsilon)+M_-(\varepsilon)\notag\\
&\hspace{0.4in}M_\uparrow(\varepsilon)+M_\downarrow(\varepsilon) \}\times\tanh(\beta\varepsilon/2),
\end{align}
with the definition ($\sigma=\uparrow,\downarrow$)
\begin{align}%\label{eq:jos}
M_{\sigma}(\varepsilon)&=[f_\sigma(-\varepsilon)]^*\partial_x f_{\sigma}(\varepsilon) - f_\sigma(\varepsilon)\partial_x[f_{\sigma}(-\varepsilon)]^*.
\end{align}
\par
\begin{figure}[h!]
\centering
\resizebox{0.48\textwidth}{!}{
\includegraphics{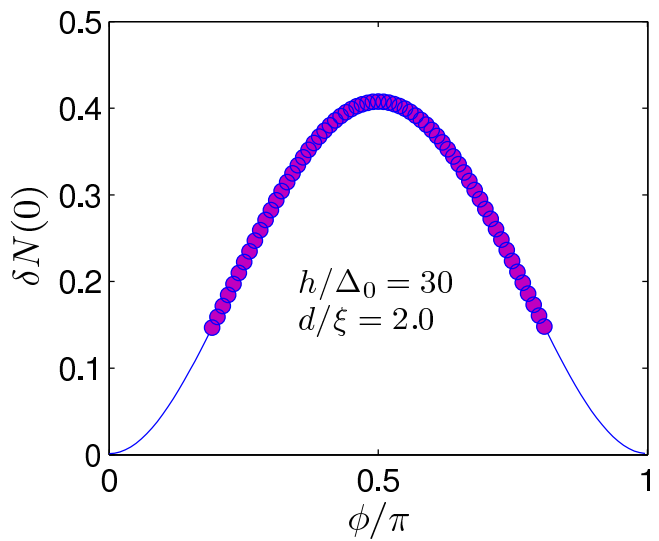}}
\caption{(Color online) Plot of the deviation from the normal-state zero-energy density of states in the middle of the ferromagnetic region $(x/d=0.5)$. The symbols denote approximately the region where the linearized treatment becomes less accurate, i.e. the anomalous Green's functions $f$ no longer satisfy $|f|\ll 1$.}
\label{fig:zep_phi}
\end{figure}

Let us first address the issue of how the zero-energy peak in the DOS treated earlier is affected by a rotation of the exchange field. In Fig. \ref{fig:zep_phi}, we plot the deviation $\delta N(0)$ from the normal-state zero-energy DOS as a function of the misorientation $\phi$ for $d/\xi=2$ and $h/\Delta_0=30$. As $\phi$ is increased, it is seen that $\delta N(0)$ grows rapidly. As it becomes comparable to the normal-state DOS in magnitude, the linearized treatment of the Usadel equations becomes less accurate, denoted by the symbols in Fig. \ref{fig:zep_phi}. 
Nevertheless, the trend seems clear: the zero-energy DOS reaches a maximum at $\phi=\pi/2$. Also, we have demonstrated 
(not shown) that the characteristics 
in Fig. \ref{fig:Fig1}b) remain the same for all $\phi$. In particular, the zero-energy peak is not destroyed by 
changing $\phi$.
\par
Examining the magnitude of the anomalous Green's function numerically, we find that the weak-proximity effect assumption $|f|\ll1$ becomes poor for energies close to zero when $\phi$ is close to $\pi/2$. Therefore, this parameter regime is strictly speaking inaccessible within our linearized treatment. Note that no such problem occurs when $\phi=0$. However, we will assume that the linearized treatment is still qualitatively correct when $\phi$ is close to $\pi/2$ in order to investigate how the critical current depends on the orientation of the exchange field orientation. 
\par
\begin{figure}[h!]
\centering
\resizebox{0.48\textwidth}{!}{
\includegraphics{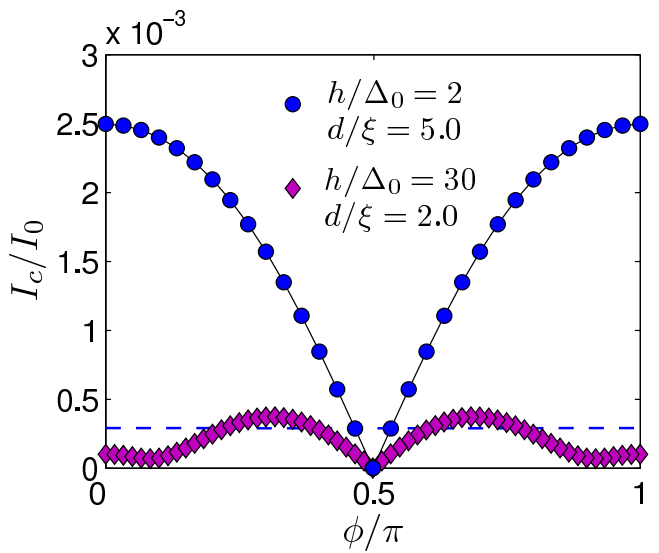}}
\caption{(Color online) Plot of the modulation of the Josephson current with orientation $\phi$ of the exchange field in the ferromagnet. The symbols correspond to an even-frequency/F/odd-frequency junction, while the dashed line denote an even-frequency/F/even-frequency junction with $h/\Delta_0=30$ and $d/\xi=2.0$.}
\label{fig:phi}
\end{figure}

In Fig. \ref{fig:phi}, we plot the variation of the critical Josephson current as a function of the orientation $\phi$ of the exchange field. If the two superconductors have a conventional even-frequency symmetry, the Josephson current is completely insensitive to the orientation of the exchange field. This is reasonable, since the superconducting order parameter in this case is spin-singlet and has no orientation in spin space. Note that a magnetic flux threading a Josephson junction in general gives rise to a Fraunhofer modulation of the current as a function of the flux. We here neglect this modification by assuming that the flux constituted by the ferromagnetic region is sufficiently weak compared to the elementary flux quantum. This is the case for either a small enough surface area or weak enough magnetization (the energy \textit{exchange splitting} may still be significant).
\par
\begin{figure}[h!]
\centering
\resizebox{0.48\textwidth}{!}{
\includegraphics{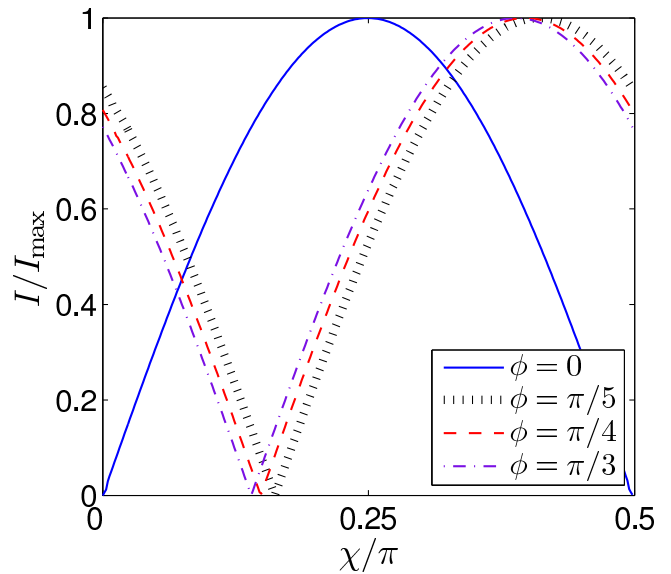}}
\caption{(Color online) Plot of the current-phase relationship for several values of $\phi$. We have chosen $h/\Delta_0=30$ and $d/\xi=2.0$.}
\label{fig:phase}
\end{figure}

The situation is quite different when one of the superconductors has an odd-frequency symmetry. In this case, the Josephson current is sensitive to the orientation $\phi$ of the exchange field, and displays the behaviour shown in Fig. \ref{fig:phi}. The reason for this is that the order parameter in the odd-frequency triplet case has a direction in spin space, here chosen as opposite-spin pairing (along the $\hat{\mathbf{z}}$-axis). For $\phi=0$, the Cooper pairs are opposite spin-paired relative to the exchange field, while for $\phi=\pi/2$ the Cooper pairs are equal-spin paired relative to the exchange field. Tuning the relative orientation of the exchange field and the superconducting order parameter in the odd-frequency superconductor is thus seen to lead to the possibility of controlling the magnitude of the Josephson current. 
In an experimental situation, only the orientation of the exchange field is probably alterable. In Fig. \ref{fig:phase}, we show the current-phase relationship for an even-frequency/F/odd-frequency Josephson junction for several values of $\phi$ to show that although it remains sinusoidal, it is shifted from $\sim \sin\phi$ to $\sim \sin[\phi+\alpha(\phi)]$ where $\alpha(\phi)$ is nonzero for $h\neq0$. 
\par
Interestingly, for $\phi=\pi/2$ the current vanishes completely as seen in Fig. \ref{fig:phi}, but it is nonzero otherwise. 
This can be understood by studying the contribution to the Josephson current from each of the components of the anomalous Green's functions. One may rewrite Eq. (\ref{eq:jos45}) as
\begin{align}
\mathbf{j} &= \mathbf{j}_t + \mathbf{j}_s + \mathbf{j}_\text{ESP},
\end{align}
where $\{j_t,j_s,j_\text{ESP}\}$ represent the contribution to the Josephson current from the $S_z=0$ triplet, the singlet, and the equal spin-pairing anomalous Green's functions, respectively. These are defined as
\begin{align}
j_t &= -M_0 \int^\infty_{-\infty} \text{d}\varepsilon \text{Re}\{2M_t(\varepsilon)\}\tanh(\beta\varepsilon/2),\notag\\
j_s &= M_0 \int^\infty_{-\infty} \text{d}\varepsilon \text{Re}\{2M_s(\varepsilon)\}\tanh(\beta\varepsilon/2),\notag\\
j_\text{ESP} &= -M_0 \int^\infty_{-\infty} \text{d}\varepsilon \text{Re}\{M_\uparrow(\varepsilon)+M_\downarrow(\varepsilon)\}\tanh(\beta\varepsilon/2),\notag\\
\end{align}
with the definition $M_0 = N_\text{F}eD/2$ and 
\begin{align}
M_j(\varepsilon)&=[f_j(-\varepsilon)]^*\partial_x f_j(\varepsilon) \notag\\
&- f_j(\varepsilon)\partial_x[f_j(-\varepsilon)]^*,\; j=\{t,s\}.
\end{align}
Now, we get $\partial _x M_t  = \partial _x (M_s  - M_\sigma  ) = 0$ $(\sigma  =  \uparrow ,\downarrow)$ for $\phi=\pi/2$ from Eq. (17). With Eq. (20), we have $M_t=0$ and similar $M_s  - M_\sigma=0$ by virtue of Eq (21) and the fact that $c_{L,R}(\varepsilon) = c_{L,R}^*(-\varepsilon)$. Therefore, the total Josephson current $j$ becomes zero for $\phi=\pi/2$. In fact, an equivalent analytical approach is viable to show that the current vanishes in the case $h=0$. In that case, one finds that $\partial_x M_t = \partial_x M_s = 0$ and $M_t=0$ at $x=0$ and $M_s=0$ at $x=d$ by means of the boundary conditions and the Usadel equation.

%We have verified this for both a large $(h\gg\Delta_0)$ and small $(h\simeq \Delta_0)$ exchange field. In the case of a large exchange-field, the vanishing of the current may be understood as follows. When $\phi=\pi/2$, it is clear from the Usadel equation that $f_t$ becomes unaffected by $h$ while $f_\sigma$ and $f_s$ are coupled and dependent on $h$. As $h$ becomes large, $f_t$ will then constitute the dominant contribution. Effectively, the Usadel equation then becomes independent of $h$, and thus the Josephson current vanishes since the current is absent for an even-frequency/N/odd-frequency junction. This qualitative argument is shown to be correct in Fig. \ref{fig:anomalous_OTE} where we plot the anomalous Green's functions as a function of $x$ at $\phi=\pi/2$. We choose $h/\Delta_0=30$ in the large exchange-field limit and $h/\Delta_0=2$ in the weak-exchange field limit, and corresponding values of $d/\xi$ to ensure a weak proximity effect. In both cases, it is seen that the singlet component $f_s$ goes to zero at the interface of the odd-frequency superconductor and the triplet component $f_t$ dominates in the large exchange-field limit, in agreement with the above argumentation. 
\par

%In Fig. \ref{fig:components}, we plot the contribution to the Josephson current (and the total current) as a function of $\phi$ for three cases: \textit{i)} no exchange field $h=0$, \textit{ii)} weak exchange field $h/\Delta_0=2$, and \textit{iii)} strong exchange field $h/\Delta_0=30$. When $h=0$, it is seen that all components vanish identically, and therefore no current may flow through the junction. When $h\neq0$, it is seen that while $I_\text{ESP}=0$ for $\phi=0$, $I_t=0$ for $\phi=\pi/2$. This is related to the orientation of the exchange field relative the quantization axis of the spin basis $\hat{\mathbf{z}}$. An interesting point seen from Fig. \ref{fig:components} is that although the critical current $I_c$ vanishes at $\phi=\pi/2$, only $I_t$ is equal to zero. The singlet contribution $I_s$ has equal magnitude but opposite sign of the equal-spin pairing contribution $I_\text{ESP}$, which leads to a cancellation of the current. This is then a different mechanism for the vanishing of the current at $h=0$, in which case all components $I_\zeta$ vanish identically.
\par
At this point, it is important to underline that although the magnitude of the Josephson current in an even-frequency/F/odd-frequency Josephson junction depends on the orientation of the exchange field, our main message is that \textit{no current} can flow in the absence of a field while the presence of an exchange field in general \textit{induces a current} except for the special case where the exchange field is parallel to the spin of the Cooper pair, $\phi=\pi/2$. 

\section{Discussion}
In our calculations, we have neglected the spatial variation of the pairing potential near the interfaces. This is permissable for either low-transmission interfaces or if the superconducting region is much less disordered than the F. \cite{bergeretRMP} Also, we have considered non-magnetic interfaces, which are routinely used in experiments. Including spin-flip scattering in the normal region is not expected to alter our qualitative conclusions since spin-flip scattering alone cannot induce triplet pairing in a normal metal in proximity to an even-frequency superconductor in the diffusive limit \cite{linder_spinflip}. Our choice of studying the diffusive limit ensures that one may disregard the generation of possible odd parity symmetry components of the superconducting gap \cite{tanakaPRL07}, which could have caused ambiguities in the interpretation of experimental results obtained in our proposed setup.
We expect that the predicted effect could be experimentally observed in disordered superconductor/ferromagnet/superconductor junctions using superconductors with different symmetries with respect to frequency. The junction widths need to be a few coherence lengths, which is well within reach with present-day technology.

\section{Summary}
In summary, we have proposed a method of identifying highly unusual superconducting states with 
the conceptually and technologically important property that the order parameter is odd under 
exchange of time-coordinates of the electrons constituting a Cooper-pair.
Remarkably, we find that an exchange field quite generally induces a Josephson effect between even- and 
odd-frequency superconductors. This constitutes a clear-cut experimental test for such an unusual
superconducting state. Since our qualitative findings rely on symmetry consideration alone, 
they are expected to be quite robust. 

\acknowledgments
The authors thank Y. Tanaka and I. B. Sperstad for useful discussions. J.L. and A.S. were supported by the Research Council of Norway, 
Grants No. 158518/432 and No. 158547/431 (NANOMAT), and Grant No. 167498/V30 (STORFORSK). 
T.Y. acknowledges support by the JSPS.

\appendix

\section{Josephson current with spin-dependent scattering}
\noindent We here provide some additional details of our calculations and also outline how spin-dependent scattering may be taken into account in the analytical expressions. We employ the linearized Usadel equations under the assumption of a weak proximity effect. This assumption is justified in the low-transparency regime (tunneling limit), where the depletion of the superconducting order parameter near the interface may also be disregarded. In the superconducting reservoirs, we employ the bulk solution which reads
\begin{align}
\hat{g}_\mathrm{SC} &= \begin{pmatrix}
c_j & 0 & 0 & s_j \e{\i\chi_j} \\
0 & c_j & \alpha_j s_j \e{\i\chi_j} & 0 \\
0 & -\alpha_j s_j \e{-\i\chi_j} & -c_j & 0 \\
-s_j\e{-\i\chi_j} & 0 & 0 & -c_j\\
\end{pmatrix},
\end{align}
where $c_j=\cosh\theta_j$, $s_j=\sinh\theta_j$, $\theta_j=\text{atanh}(\Delta_j(\varepsilon)/\varepsilon)$, and $j = L,R$ denotes the left and right superconducting region. Here, $\chi_j$ denotes the broken U(1) phase in superconductor $j$, and we use the convention $\chi_R = -\chi_L \equiv \chi$. For an odd-frequency superconductor on side $j$, we have $\alpha_j = 1$ while for an even-frequency superconductor on side $j$, one has $\alpha_j=-1$. 
\par
The Kupriyanov-Lukichev \cite{kupluk} boundary conditions now read
\begin{align}
2\gamma d \hat{g}\partial_x \hat{g} &= -[\hat{g}, \hat{g}_L]|_{x=0},\notag\\
2\gamma d \hat{g}\partial_x \hat{g} &= [\hat{g}, \hat{g}_R]|_{x=d}.
\end{align}
In the normal region, we may write the Green's function as
\begin{align}
\hat{g}^\text{R} = \begin{pmatrix}
\underline{1} & f^\text{R}_\text{t}(\varepsilon)\underline{\tau_1} +  f^\text{R}_\text{s}(\varepsilon)\i\underline{\tau_2}\\
-[f^\text{R}_\text{t}(-\varepsilon)\underline{\tau_1}+f^\text{R}_\text{s}(-\varepsilon)\i\underline{\tau_2}]^* & -\underline{1}\\
\end{pmatrix},
\end{align} 
where the subscripts 't' and 's' denote the triplet and singlet part of the anomalous Green's function. Note that the triplet part is odd in frequency. In the following, we will consider an exchange field $\mathbf{h} \parallel \hat{\mathbf{z}}$, i.e. perpendicular to the spin of the Cooper pair, such that there are no equal-spin pairing components $f_\sigma$ ($\sigma=\uparrow,\downarrow$) of the anomalous Green's function. Introducing $f_\pm = f_t \pm f_s$, we may write the boundary conditions more explicitely. At $x=0$ one obtains
\begin{align}\label{eq:bc1}
2\gamma d \partial_x f_+ &= 2f_+ c_L - 2s_L \e{-\i\chi},\notag\\
2\gamma d \partial_x f_- &= 2f_- c_L - 2\alpha_Ls_L \e{-\i\chi},
\end{align}
while the same procedure at $x=d$ yields
\begin{align}\label{eq:bc2}
2\gamma d \partial_x f_+ &= 2s_R \e{\i\chi} - 2f_+ c_R,\notag\\
2\gamma d \partial_x f_- &= 2\alpha_Rs_R \e{\i\chi} - 2f_-c_R.
\end{align}
The linearized Usadel equations in the normal region may be formally obtained by assuming that $|f_\pm|\ll1$: \cite{Demler,Houzet,linder}
\begin{align}
\partial_x^2 f_t \pm \partial_x^2 f_s + A_\pm f_t \pm B_\pm f_s = 0,
\end{align}
where we have introduced
\begin{align}
A_\pm &= \frac{1}{D}\Big[2\i(\varepsilon\pm h) - g_\text{so} - \frac{g_\text{sf}S_z}{2}\Big],\notag\\
B_\pm &= \frac{1}{D}\Big[2\i(\varepsilon\pm h) - \frac{g_\text{sf}(2S_{xy} + S_z)}{2}\Big].
\end{align}
It is possible to find a general analytical solution for the functions $\{f_t,f_s\}$, and this can be written as
\begin{align}
f_t &= c_1 \e{-k_-x} + c_2\e{k_-x} + c_3\e{-k_+x} + c_4\e{k_+x},\notag\\
f_s &= \frac{c_1S_- \e{-k_-x} + c_2S_-\e{k_-x} + c_3S_+\e{-k_+x} + c_4S_+\e{k_+x}}{2(B_--B_+)},
\end{align}
where $\{c_i\}$ are constants to be determined from the boundary conditions Eq. (\ref{eq:bc1}) and (\ref{eq:bc2}).
Also, we have defined the auxiliary quantities:
\begin{align}
k_\pm &= \frac{1}{2}\sqrt{(-A_+-A_--B_+-B_-)\pm R},\notag\\
S_\pm &= (A_++A_--B_+-B_-)\pm R,\notag\\
R &= [(A_++A_-)^2+(B_++B_-)^2 - 4(A_+B_-+A_-B_+) \notag\\
&\;\;\;\; + 2(A_+-A_-)(B_+-B_-) ]^{1/2}.
\end{align}
First, we note that
\begin{align}
f_\pm &= c_1\e{-k_-x}(1\pm L_-) + c_2\e{k_-x}(1\pm L_-) \notag\\
&+ c_3\e{-k_+x}(1\pm L_+) + c_4\e{k_+x}(1\pm L_+),
\end{align}
with the definition $L_\pm = \i DS_\pm/(8h)$. We also introduce
\begin{eqnarray}
\delta_{L\pm} = 
\begin{cases}
1        & \text{for } + \\
\alpha_L &  \text{for } -\\
\end{cases} 
\end{eqnarray}
and similarly for $L\to R$.
After lengthy calculations, we finally arrive at an explicit expression for the coefficients $\{c_i\}$:
\begin{widetext}
\begin{align}
c_4 &= \frac{G_+E_-/E_+ - G_-}{F_- - F_+E_-/E_+},\; c_3 = -(G_++c_4F_+)/E_+,\; c_2 = Y_0 + Y_3c_3 + Y_4c_4,\; c_1 = X_0 + X_2c_2 + X_3c_3 + X_4c_4.
\end{align}
We have defined the auxiliary quantities:
\begin{align}
P &= \frac{1-L_-}{1+L_-}, \; X_0 = \frac{\delta_{L+}s_L\e{-\i\chi}}{(1+L_-)(c_L + \gamma dk_-)},\; X_2 = \frac{\gamma dk_- -  c_L}{\gamma d k_- + c_L},\notag\\
X_3 &= - \frac{(1+L_+)(c_L + \gamma dk_+)}{(1+L_-)(c_L + \gamma dk_-)}, \; X_4 = \frac{(1+L_+)(\gamma dk_+ - c_L)}{(1+L_-)(\gamma dk_- + c_L)},\notag\\
\text{ }\\
Y_0 &= \frac{s_L\e{-\i\chi}(P\delta_{L+}-\delta_{L-})}{(c_L-\gamma dk_-)[P(1+L_-)-(1-L_-)]},\notag\\
Y_3 &= \frac{(1-L_+)(c_L + \gamma d k_+) - (1+L_+)(Pc_L + P\gamma d k_+)}{(c_L-\gamma dk_-)[P(1+L_-)-(1-L_-)]},\notag\\
Y_4 &= \frac{(1-L_+)(c_L - \gamma d k_+) - (1+L_+)(Pc_L - P\gamma d k_+)}{(c_L-\gamma dk_-)[P(1+L_-)-(1-L_-)]},\notag\\
\text{ }\\
E_\pm &= 2\e{-k_+d}(1\pm L_+)(c_R-\gamma dk_+) + 2(1\pm L_-)[Y_3\e{k_-d}(c_R+\gamma dk_-) + (X_2Y_3+X_3)\e{-k_-d}(c_R-\gamma dk_-)],\notag\\
F_\pm &= 2\e{k_+d}(1\pm L_+)(c_R+\gamma dk_+) + 2(1\pm L_-)[Y_4\e{k_-d}(c_R+\gamma dk_-) + (X_2Y_4+X_4)\e{-k_-d}(c_R-\gamma dk_-)],\notag\\
G_\pm &= -2\delta_{R\pm}s_R\e{\i\chi} + 2(1\pm L_-)[Y_0\e{k_- d}(c_R+\gamma dk_-) + (X_2Y_0 + X_0)\e{-k_- d}(c_R-\gamma dk_-)].
\end{align}
The above equations may be considerably simplified by considering only uniaxial spin-flip scattering. Setting the planar spin-flip and spin-orbit scattering rates to zero, we obtain the anomalous Green's function as
\begin{align}
f_\pm = c_{1\pm}\e{\i k_\pm x} + c_{2\pm}\e{-\i k_\pm x},\; k_\pm = \sqrt{\frac{4\i(\varepsilon\pm h) - 3g_\text{sf}}{2D}},
\end{align}
with the coefficients $(\rho_\sigma=\i\gamma dk_\sigma,\; \sigma=\pm)$
\begin{align}
c_{2\sigma} = \frac{\delta_{L\sigma}s_L\e{-\i\chi}\e{\i k_\sigma d}(\rho_\sigma+c_R) + \delta_{R\sigma}s_R\e{\i\chi}(\rho_\sigma-c_L)}{\e{\i k_\sigma d}(\rho_\sigma +c_L)(\rho_\sigma +c_R) - \e{-\i k_\sigma d}(\rho_\sigma -c_R)(\rho_\sigma -c_L)},\; c_{1\sigma} &= \frac{c_{2\sigma}(\rho_\sigma+c_L) - \delta_{L\sigma}s_L\e{-\i\chi}}{\rho_\sigma-c_L},\; \sigma=\pm.
\end{align}
Once $\hat{g}^\text{R}$ has been obtained, the Josephson current may be calculated according to the formulas in the main text.

\section{Pauli matrices}
\noindent The Pauli-matrices used in this paper are defined as 
\begin{align}
\underline{\tau_1} &= \begin{pmatrix}
0 & 1\\
1 & 0\\
\end{pmatrix},\;
\underline{\tau_2} = \begin{pmatrix}
0 & -\i\\
\i & 0\\
\end{pmatrix},\;
\underline{\tau_3} = \begin{pmatrix}
1& 0\\
0& -1\\
\end{pmatrix},\notag\\
\underline{1} &= \begin{pmatrix}
1 & 0\\
0 & 1\\
\end{pmatrix},\;
\hat{1} = \begin{pmatrix}
\underline{1} & \underline{0} \\
\underline{0} & \underline{1} \\
\end{pmatrix},\;
\hat{\tau}_i = \begin{pmatrix}
\underline{\tau_i} & \underline{0}\\
\underline{0} & \underline{\tau_i} \\
\end{pmatrix},\notag\\
\hat{\rho}_1 &= \begin{pmatrix}
\underline{0} & \underline{\tau_1}\\
\underline{\tau_1} & \underline{0} \\
\end{pmatrix},\;
\hat{\rho}_2 =  \begin{pmatrix}
\underline{0} & -\i\underline{\tau_1}\\
\i\underline{\tau_1} & \underline{0} \\
\end{pmatrix},\;
\hat{\rho}_3 = \begin{pmatrix}
\underline{1} & \underline{0}\\
\underline{0} & -\underline{1}  \\
\end{pmatrix}.
\end{align}

\end{widetext}

\end{document}